\documentclass[preprint]{revtex4}
\usepackage{amsmath}

\begin{document}

\title{Truncated Wigner approximation as a non-positive Kraus map}
\author{A.B. Klimov, I. Sainz, J.L. Romero}
\date{\today}

\begin{abstract}
We show that the Truncated Wigner Approximation developed in the flat
phase-space is mapped into a Lindblad-type evolution with an indefinite
metric in the space of linear operators. As a result, the classically
evolved Wigner function corresponds to a non-positive operator $\hat{R}(t)$,
which does not describe a physical state. The rate of appearance of negative
eigenvalues of $\hat{R}(t)$ can be efficiently estimated. The short-time
dynamics of the Kerr and second harmonic generation Hamiltonains are
discussed.
\end{abstract}

\maketitle

\address{ Dept. de F\'{\i}sica, Universidad de Guadalajara, 44420
Guadalajara, Mexico\\}


\section{Introduction}

The Liouvillian, or Truncated Wigner Approximation (TWA) is one of the most
popular semiclassical approximations, which has been widely used in numerous
applications (see \cite{polkovnikov} for a recent review). Developed in the
framework of the phase-space approach \cite{PS}, \cite{berry}, the TWA
allows us to employ classical intuition in order to describe the initial
stage of the quantum evolution.

According to the general ideas of the phase-space mapping, every operator $%
\hat{f}$ acting in the Hilbert space of a quantum system is in one-to-one
correspondence with its symbol $W_{f}(\Omega )$, defined in the classical
phase-space $\mathcal{M}$, $\Omega \in \mathcal{M}$,%
\begin{equation}
\hat{f}\Leftrightarrow W_{f}(\Omega )  \label{map}
\end{equation}%
This procedure allows the computation of average values of any $\hat{f}$ \
by convoluting $W_{f}(\Omega )$ with the symbol of the density matrix $%
W_{\rho }(\Omega )$ (the Wigner function). The von Neumann evolution
equation for the density matrix is mapped into the Moyal equation \cite%
{Moyal},

\begin{equation}
\partial _{t}W_{\rho }(\Omega |t)=\{W_{\rho }(\Omega |t),W_{H}(\Omega
)\}_{M},\;  \label{ME}
\end{equation}%
where $W_{H}$ is the symbol of the Hamiltonian and $\{,\}_{M}$ denote the
Moyal brackets. Equation (\ref{ME}) in general contains higher order
derivatives on the phase-space coordinates, which makes its solution a
difficult task. Nevertheless, Eq.(\ref{ME}) admits an expansion in powers of
a semiclassical parameter $\epsilon \ll 1$, and acquires the Liouvillian
form,
\begin{equation}
\partial _{t}W_{\rho }(\Omega )\approx \epsilon \{W_{\rho }(\Omega
),W_{H}(\Omega )\}_{P},  \label{twa eq}
\end{equation}%
to leading order in $\epsilon $, where $\{,\}_{P}$ is the Poisson bracket on
$\mathcal{M}$. The solution of Eq.(\ref{twa eq}), that approximates the
exact Wigner function $W_{\rho }(\Omega |t)$ as a ``classically evolve'',%
\begin{equation*}
W_{\rho }(\Omega |t)\approx W_{\rho }(\Omega ^{cl}(-t)),
\end{equation*}%
where $\Omega ^{cl}(t)$ denotes classical trajectories, is known as the
Truncated Wigner Approximation (TWA). This approximation describes the
propagation of every point of the initial distribution along the
corresponding classical trajectory. The evolution of average values are
computed by integrating symbols of operators with $W_{\rho }(\Omega
^{cl}(-t))$,
\begin{equation}
\langle \hat{f}(t)\rangle \approx \int d\Omega W_{f}(\Omega )W_{\rho
}(\Omega ^{cl}(-t)).  \label{avarage2}
\end{equation}

Initially applied to the analysis of semiclassical dynamics of quantum
systems in the flat $p-q$ phase space, \cite{heller}, \cite{drobny} (see
also \cite{polkovnikov}, \cite{Miller} and references therein), TWA was
extended to quantum systems with SU(2) \cite{su2evol}, \cite{rev} and SU(3)
symmetries \cite{rev}, \cite{suN}.\ TWA leads to exact results only for
harmonic quantum dynamics. In the flat $p-q$\ phase-space a harmonic
evolution is governed by at most quadratic (in the phase-space variables)
Hamiltonians, and leads to a symplectic deformation of the initial
distribution (interestingly, a similar behavior is also observed for some
dissipative scenarios \cite{OdeA}). For quantum systems with a semi-simple
dynamic symmetry group the harmonic evolution is generated by Hamiltonians
linear in the group generators. In this case the initial distribution is
rigidly displaced (i.e. without distortion) in the corresponding phase-space
as a consequence of the covariance of phase-space distributions under group
transformations.

Although formally speaking, TWA fails from the very beginning in case of
non-linear evolution \cite{ole} (since the classical dynamics preserves the
phase-space area,), it describes relatively well a short-time non-linear
dynamics of smooth and localized distributions (representing the so-called
semiclassical states). In the simplest case of quantum systems with the
Heisenberg-Weyl symmetry, it was noted \cite{bracken} by using the Hudson
theorem that the operator $\hat{R}(t)$ corresponding to the inverse map of $%
W_{\rho }(\Omega ^{cl}(-t))$, in general, does not describe a physical state.

In the present paper we will analyze the evolution equation for $\hat{R}(t)$
and show that it has an indefinite Lindblad form, i.e. contains positive and
negative Lindblad-like terms. In other words, an anharmonic classical
dynamics, if viewed from the quantum point of view, corresponds to a very
specific\textit{\ non-unitary} quantum evolution which results in a
non-positivity of $\hat{R}(t)$. We discuss the algebraic structure of this
equation and the physical implications of its non-positivity. We show that
negative eigenvalues of the operator $\hat{R}(t)$ appear already at $t=0^{+}$
with the rate that depends both on the initial state and the degree of
non-linearity of the Hamiltonian.

Here we will focus only on the evolution in the flat phase-space, although a
similar approach can be applied to quantum systems with higher symmetries.

\section{Operator form of the Liouvillian evolution equation}

\subsection{General considerations}

In case of the Heisenberg-Weyl symmetry the map from the Hilbert space to
the flat phase-space is defined by the kernel
\begin{eqnarray}
\hat{w}(\alpha ) &=&\hat{D}(\alpha )(-1)^{\hat{a}^{\dagger }\hat{a}}\hat{D}%
^{\dagger }(\alpha ),  \label{w} \\
\mathrm{Tr}\hat{w}(\alpha ) &=&1,\;\int \frac{d^{2}\alpha }{\pi }\hat{w}%
(\alpha )=\delta ^{2}(\alpha ),  \notag
\end{eqnarray}%
in such a way that \cite{wf}
\begin{equation}
W_{f}(\alpha )=\mathrm{Tr}\left( \hat{f}\,\hat{w}(\alpha )\right) ,
\label{map gen}
\end{equation}%
and the inverse transformation has the form%
\begin{equation}
\hat{f}=\int \frac{d^{2}\alpha }{\pi }\hat{w}(\alpha )W_{f}(\alpha ),
\label{rec gen}
\end{equation}%
where $\hat{D}(\alpha )=\exp (\alpha \hat{a}^{\dagger }-\alpha ^{\ast }\hat{a%
})$, is the displacement operator and $d^{2}\alpha =d\alpha d\alpha ^{\ast }$%
. The Wigner function automatically satisfies the normalization condition,%
\begin{equation}
\int \frac{d^{2}\alpha }{\pi }W_{\rho }(\alpha (-t))=1,  \label{NW}
\end{equation}%
corresponding to $\mathrm{Tr}\hat{\rho}=1$. The purity of the state, $P=%
\mathrm{Tr}\hat{\rho}^{2}$, is expressed in terms of the Wigner function as%
\begin{equation}
P=\int \frac{d^{2}\alpha }{\pi }W_{\rho }^{2}(\alpha ).  \label{P}
\end{equation}

The TWA evolution equation (\ref{twa eq}) takes the form%
\begin{equation}
i\partial _{t}W_{\rho }=\partial _{\alpha }W_{H}\partial _{\alpha ^{\ast
}}W_{\rho }-\partial _{\alpha ^{\ast }}W_{H}\partial _{\alpha }W_{\rho
}=i\{W_{H},W_{\rho }\}_{P},  \label{twa eq1}
\end{equation}%
leading to the Liouvillian evolution of the Wigner function
\begin{equation}
W_{\rho }(\alpha )\rightarrow W_{\rho }(\alpha |t)\approx W_{\rho }(\alpha
(-t)),  \label{Wt}
\end{equation}%
where the canonical transformation $\alpha \rightarrow \alpha (t)$,
generated by $W_{H}(\alpha )$, is described by the classical Hamilton
equations%
\begin{equation*}
\dot{\alpha}(t)=-i\partial _{\alpha ^{\ast }}W_{H}(\alpha ).
\end{equation*}%
Eq.(\ref{Wt}) contradicts Hudson%
\'{}%
s Theorem \cite{hudson} in case of anharmonic dynamics (when $W_{H}(\alpha )$
contains grater than quadratic powers of $\alpha $ and $\alpha ^{\ast }$).
This is clearly seen in the example of evolution of the Wigner function of
the initial coherent state, $|\alpha _{0}\rangle $,%
\begin{equation}
W_{\rho }(\alpha )=2\exp (-|\alpha -\alpha _{0}|^{2}),  \label{WG}
\end{equation}%
which evolves according to (\ref{Wt}) into a positive, but non-Gaussian
function, thus not corresponding to any physical quantum state \cite{bracken}%
.

The inverse map of the evolved Wigner function $W_{\rho }(\alpha (-t))$,
leads to the following time-dependent Hermitian operator%
\begin{eqnarray}
\hat{R}(t) &=&\int \frac{d^{2}\alpha }{\pi }\hat{w}(\alpha )W_{\rho }(\alpha
(-t)),  \label{R} \\
\hat{R}(0) &=&\hat{\rho}.  \notag
\end{eqnarray}%
The average values calculation according to Eq.(\ref{avarage2}) is
equivalent to tracing $\hat{R}(t)$ with the corresponding operator
\begin{equation*}
\langle \hat{f}(t)\rangle \approx Tr\left( \hat{R}(t)\hat{f}\right) .
\end{equation*}

Since the classical evolution is reduced to the canonical transformation of
the phase-space coordinates the function $W_{\rho }(\alpha (-t))$ satisfies
the normalization condition (\ref{NW}) and the purity (\ref{P})
\textquotedblleft conservation\textquotedblright
\begin{equation}
P(t)=\int \frac{d^{2}\alpha }{\pi }W_{\rho }^{2}(\alpha (-t))=Tr\hat{R}%
^{2}(t)=P(0).  \label{Pt}
\end{equation}%
Thus, for the initial pure state, the operator (\ref{R}) fulfills the
conditions
\begin{eqnarray}
Tr\hat{R}(t) &=&1,  \label{R1} \\
Tr\hat{R}^{2}(t) &=&1,  \label{R2}
\end{eqnarray}%
which, however, does not mean that $\hat{R}^{2}(t)$ is equal to $\hat{R}(t)$%
, except in the case when the Hamiltonian is a quadratic function of $a$ and
$a^{\dagger }$, as shown in Appendix A. Moreover, the trace-class operator $%
\hat{R}(t)$ does not describe a physical state (except for the harmonic
evolution), which is reflected in appearance of negative eigenvalues for $%
t=0^{+}$ as shown below.

\subsection{Evolution equation for $\hat{R}(t)$}

It is instructive to analyze the non-positivity of $\hat{R}(t)$ operator on
the level of the evolution equation in the Hilbert space.

Let us consider a symmetrized $n+m$ degree Hamiltonian%
\begin{equation}
\hat{H}_{mn}=\{\hat{a}^{\dagger m}\hat{a}^{n}+\hat{a}^{\dagger n}\hat{a}%
^{m}\}_{sym},\;n\geq m,  \label{Hq}
\end{equation}%
where $\{..\}_{sym}$ means the full normalized symmetrization of the
monomial $\hat{a}^{\dagger n}\hat{a}^{m}$, see Appendix C. The (symmetrized)
monomial Hamiltonian describes a variety of physical processes \cite{heller}%
. In addition, an arbitrary Hamiltonian on $a$ and $a^{\dagger }$ can be
represented as a series on $\hat{H}_{mn}$ \cite{H}.

The symbol of (\ref{Hq}) is%
\begin{equation*}
W_{H}(\alpha ,\alpha ^{\ast })=\alpha ^{\ast m}\alpha ^{n}+\alpha ^{m}\alpha
^{\ast n}.
\end{equation*}%
and the Liouville equation (\ref{twa eq1}) takes the form%
\begin{equation}
i\partial _{t}W_{\rho }=\alpha ^{\ast m-1}\alpha ^{n-1}\left( n\alpha ^{\ast
}\partial _{\alpha ^{\ast }}W_{\rho }-m\alpha \partial _{\alpha }W_{\rho
}\right) +\alpha ^{m-1}\alpha ^{\ast n-1}\left( m\alpha ^{\ast }\partial
_{\alpha ^{\ast }}W_{\rho }-n\alpha \partial _{\alpha }W_{\rho }\right) .
\label{EqW}
\end{equation}%
It can be shown (see Appendix B) that for the generic Hamiltonian (\ref{Hq})
the equation for $\hat{R}(t)$ defined in (\ref{R}) can be reduced to the
Lindblad-type \cite{lindblad} form%
\begin{equation}
\partial _{t}\hat{R}=i[\hat{R},\hat{H}_{eff}]+\mathcal{L}(\hat{R}),
\label{Req}
\end{equation}%
where the effective Hamiltonian is
\begin{equation}
\hat{H}_{eff}=\frac{n+m}{2^{n+m}}\left( \hat{a}^{\dagger m}\hat{a}^{n}+\hat{a%
}^{\dagger n}\hat{a}^{m}+\hat{a}^{n}\hat{a}^{\dagger m}+\hat{a}^{m}\hat{a}%
^{\dagger n}\right) ,  \label{Heff}
\end{equation}%
and the $\mathcal{L}$ has the following structure%
\begin{equation}
\mathcal{L}=\mathcal{G}_{nm}^{(0)}+\mathcal{G}_{mn}^{(1)}+\mathcal{G}%
_{nm}^{(1)}+\mathcal{G}_{mn}^{(2)}+\mathcal{G}_{nm}^{(2)},  \label{lin}
\end{equation}%
where
\begin{eqnarray}
\mathcal{G}_{nm}^{(0)} &=&\frac{n-m}{2^{n+m+1}}\left( \mathcal{L}_{0m}^{nm}+%
\tilde{\mathcal{L}}_{n0}^{nm}-\mathcal{L}_{n0}^{nm}-\tilde{\mathcal{L}}%
_{0m}^{nm}\right) ,  \label{lin1} \\
\mathcal{G}_{mn}^{(1)} &=&\frac{n}{2^{n+m+1}}\sum_{j=1}^{m-1}C_{m}^{j}\left(
\mathcal{L}_{j0}^{mn}+\tilde{\mathcal{L}}_{jn}^{mn}-\mathcal{L}_{jn}^{mn}-%
\tilde{\mathcal{L}}_{j0}^{mn}\right) ,  \label{lin2} \\
\mathcal{G}_{mn}^{(2)} &=&\frac{1}{2^{n+m+1}}\sum_{j=0}^{m}%
\sum_{k=1}^{[n/2]}C_{m}^{j}C_{n}^{k}(n-2k)\left( \mathcal{L}_{jk}^{mn}+%
\tilde{\mathcal{L}}_{jk}^{mn}-\mathcal{L}_{jk}^{mn}-\tilde{\mathcal{L}}%
_{jk}^{mn}\right) ,  \label{lin3}
\end{eqnarray}%
where operators $L_{kj}^{pq}$ and $\tilde{L}_{kj}^{pq}$ defining the
Lindblad super-operators
\begin{eqnarray}
\mathcal{L} &=&2L\otimes L^{\dagger }-L^{\dagger }L\otimes \hat{I}-\hat{I}%
\otimes L^{\dagger }L,  \label{LL1} \\
\tilde{\mathcal{L}} &=&2\tilde{L}\otimes \tilde{L}^{\dagger }-\tilde{L}%
^{\dagger }\tilde{L}\otimes \hat{I}-\hat{I}\otimes \tilde{L}^{\dagger }%
\tilde{L},  \label{LL2}
\end{eqnarray}%
are%
\begin{eqnarray*}
L_{jk}^{pq} &=&\hat{a}^{\dagger j}\hat{a}^{k}-i\hat{a}^{p-j}\hat{a}^{\dagger
q-k},\;\tilde{L}_{jk}^{pq}=(L_{kj}^{qp})^{\dag }, \\
p,q &=&m,n;\;j=0,..p;\;k=0,...,q,
\end{eqnarray*}%
and $C_{m}^{j}$ are the binomial coefficients.

The following observations about the structure of Eq.(\ref{Req}-\ref{lin3})
can be made:

\begin{description}
\item[1.] $\mathcal{L}(\hat{R})=0$ and $\hat{H}_{eff}=\hat{H}_{mn}$ in case
of harmonic evolution, i.e. when the Hamiltonian is a quadratic form on
quadratic function of $a$ and $a^{\dagger }$.

\item[2.] The evolution of $\hat{R}(t)$\ is not unitary for $n+m>2$, in the
sense that $\hat{R}(t)\neq \hat{R}^{2}(t)$\ for $t>0$\ (see Appendix B).
This, nevertheless, does not mean that a pure state decoheres into a mixed
state as it evolves according to Eq. (\ref{Req}). It is discussed below that
this specific non-unitarity along with the condition (\ref{Pt}) leads to the
appearance of negative eigenvalues of the operator $\hat{R}(t)$.

\item[3.] The Hamiltonian \textquotedblleft sector\textquotedblright\ of the
evolution decreases for higher non-linearities. In addition, $\hat{H}%
_{eff}\sim \hat{H}_{mn}$ only for $m=1$ and $\hat{H}_{eff}\sim \frac{1}{2}%
\hat{H}_{22}+\frac{1}{2}$ in the particular case $n=m$ $=2$ (see Appendix C).

\item[4.] The number of positive and negative coefficients of the Lindblad
operators, $\mathcal{L}_{kj}^{nm},\mathcal{\tilde{L}}_{kj}^{nm}\geq 0$, is
the same. The evolution of $\hat{R}(t)$ is induced by the Kraus map \cite%
{kraus} $\hat{\rho}\rightarrow S(\hat{\rho})$ of the form%
\begin{equation}
S=\hat{K}_{0}\otimes \hat{K}_{0}^{\dagger }+\sum_{j}\left( \hat{K}%
_{+j}\otimes \hat{K}_{+j}^{\dagger }-\hat{K}_{-j}\otimes \hat{K}%
_{-j}^{\dagger }\right) ,  \label{K}
\end{equation}%
\begin{equation*}
\hat{K}_{0}\hat{K}_{0}^{\dagger }+\sum_{j}\left( \hat{K}_{+j}\hat{K}%
_{+j}^{\dagger }-\hat{K}_{-j}\hat{K}_{-j}^{\dagger }\right) =\hat{I}.
\end{equation*}%
Thus, the map generated by the classical dynamics is not completely positive
\cite{maps}. It will be shown below that such a map is actually \textit{%
non-positive}.

\item[5.] In general, the non-Hamiltonian part of the evolution equation (%
\ref{Req}) has the structure:
\begin{equation}
\mathcal{L}(\hat{R})=i\left( F(\hat{R})-F^{\dagger }(\hat{R})\right) ,
\label{LF}
\end{equation}%
where
\begin{equation*}
F=\sum_{j}\hat{A}_{j}\otimes \hat{B}_{j},
\end{equation*}%
is a (non-Lindbladian) map, which leads to the conservation conditions (\ref%
{R1}, \ref{R2}) and, as a consequence, to the following overlap relation%
\begin{equation}
Tr\left( \mathcal{L}(\hat{R}(t))\hat{R}(t)\right) =0.  \label{OR1}
\end{equation}

\item[6.] For an initial pure state it follows from (\ref{R1} - \ref{R2})
that
\begin{equation*}
\sum_{k}\lambda _{k}(t)=1,~~\sum_{k}\lambda _{k}^{2}(t)=1,
\end{equation*}%
and thus
\begin{equation}
\sum_{k}\lambda _{k}(t)\left( 1-\lambda _{k}(t)\right) =0,  \label{ll}
\end{equation}%
where $\{\lambda _{k}(t),k=1,2,..\}$ are eigenvalues of the $\hat{R}(t)$
operator \cite{papa}%
\begin{equation}
\hat{R}(t)=\sum_{k}\lambda _{k}(t)|k(t)\rangle \langle k(t)|,\quad \hat{R}%
(0)=|\psi _{0}\rangle \langle \psi _{0}|.  \label{Rsm}
\end{equation}%
According to Eq.(\ref{RR}) the rank of the operator $\hat{R}(t)$ is not
preserved by the equation of motion (\ref{Req}). Thus, the initial (first
rank) density matrix evolves into the form (\ref{Rsm}) in the anharmonic
case, since $|\lambda _{k}(t>0)|<1$, unless the initial states is an
eigenstate of the Hamiltonian (\ref{Hq}),
\begin{equation}
\lbrack \hat{R}(0),\hat{H}]=0.  \label{LR0}
\end{equation}%
It immediately follows from the relation (\ref{ll}) that at least one
negative eigenvalue of $\hat{R}(t)$ appears at $t=0^{+}$ for an initial pure
state if the condition (\ref{LR0}) is not fulfilled. The rates of appearance
of negative eigenvalues depend both on the Hamiltonian and the initial state.
\end{description}

The upper bound of negative eigenvalues can be estimated by using the
min-max theorem \cite{papa} in the subspace orthogonal to the initial state $%
|\psi _{0}\rangle $, i.e. finding a state $|\phi \rangle $, $\langle \psi
_{0}|\phi \rangle =0$, such that
\begin{equation}
\lambda _{min}\leq \min_{\phi }\langle \phi |\hat{R}(t)|\phi \rangle <0.
\label{lmin1}
\end{equation}%
For short times, when%
\begin{equation}
\hat{R}(t)\approx \hat{R}_{0}+it[\hat{R}_{0},\hat{H}_{eff}]+t\mathcal{L}(%
\hat{R}_{0}),  \label{stE}
\end{equation}%
the condition Eq.(\ref{lmin1}) is reduced to
\begin{equation}
\lambda _{min}\leq t\min_{\phi }\langle \phi |\mathcal{L}(R_{0})|\phi
\rangle <0.  \label{min}
\end{equation}

It is worth noting that the evolution equation in the form (\ref{Req}-\ref%
{lin3}) greatly simplifies the estimation of the self-correlation function
\begin{equation}
G(t)=\langle \psi (0)|\hat{R}(t)|\psi (0)\rangle ,  \label{Gt}
\end{equation}%
and the fidelity
\begin{equation}
\mathcal{F}(t)=\langle \psi (t)|\hat{R}(t)|\psi (t)\rangle ,  \label{Fg}
\end{equation}%
where $|\psi (t)\rangle $ is the exact state vector. The deviation of $%
\mathcal{F}(t)$ from the identity, in particular, in the beginning of
evolution, quantifies the quality of TWA.

\section{Examples}

In this Section we consider two representative examples of anharmonic
evolution.

\subsection{Kerr evolution}

Kerr dynamics is generated by the following (symmetrized) fourth-order
Hamiltonian%
\begin{equation*}
\hat{H}_{Kerr}=\{2\hat{a}^{\dagger 2}\hat{a}^{2}\}_{sym}.
\end{equation*}%
The effective Hamiltonian (\ref{Heff}) is
\begin{equation}
\hat{H}_{eff}=\frac{1}{2}\left( \hat{a}^{\dagger 2}\hat{a}^{2}+\hat{a}^{2}%
\hat{a}^{\dagger 2}\right) =\frac{1}{2}\hat{H}_{K}+\frac{1}{2},
\label{HeffK}
\end{equation}%
and the non-unitary part of the evolution equation (\ref{lin}-\ref{lin3})
takes the form%
\begin{equation}
\mathcal{L}_{Kerr}=\frac{1}{4}\left( \mathcal{L}_{10}^{22}+\tilde{\mathcal{L}%
}_{12}^{22}-\mathcal{L}_{12}^{22}-\tilde{\mathcal{L}}_{10}^{22}\right) ,
\label{LK}
\end{equation}%
where the operators defining (\ref{LL1})-(\ref{LL2}) are%
\begin{align*}
L_{10}^{22} &=\hat{a}^{\dagger }-i\hat{a}\hat{a}^{\dagger 2}, & \tilde{L}%
_{10}^{22}&=\hat{a}^{\dagger }+i\hat{a}\hat{a}^{\dagger 2}, \\
\tilde{L}_{12}^{22}&=i\left(\hat{a}-i\hat{a}^{\dagger}\hat{a}^{ 2}\right) ,
& L_{12}^{22} &=-i\left(\hat{a}+i\hat{a}^{\dagger }\hat{a}^{2}\right).
\end{align*}%
It is instructive to represent (\ref{LK}) in form (\ref{LF}) where%
\begin{equation}
F(\hat{R})=\frac{1}{2}\left( \hat{a}\hat{R}\hat{a}^{\dagger 2}\hat{a}+\hat{a}%
^{\dagger }\hat{R}\hat{a}^{2}\hat{a}^{\dagger }\right) .  \label{FRk}
\end{equation}%
The Kraus map $\hat{R}(t)\rightarrow \hat{R}(t+\delta t)$ has the form (\ref%
{K}) with
\begin{eqnarray*}
\hat{K}_{+1} &=&\frac{1}{4}\sqrt{\delta t}L_{10}^{22},\;\hat{K}_{+2}=\frac{1%
}{4}\sqrt{\delta t}\;\tilde{L}_{12}^{22}, \\
\hat{K}_{-1} &=&\frac{1}{4}\sqrt{\delta t}\tilde{L}_{10}^{22},\;\hat{K}_{-2}=%
\frac{1}{4}\sqrt{\delta t}\;L_{12}^{22}, \\
\hat{K}_{0} &=&\hat{I}-i\delta t\hat{H}_{eff}
\end{eqnarray*}%
where $H_{eff}$ is defined in (\ref{HeffK}).

In what follows we analyze evolution of negative eigenvalues of $\hat{R}(t)$.

\begin{description}
\item[a)] Fock states $|n\rangle $ do not evolve under action of Kerr
Hamiltonian, $[|n\rangle \langle n|,H_{Kerr}]=0$. It is straightforward to
see that $[|n\rangle \langle n|,H_{eff}]=0$ and $\mathcal{L}%
_{Kerr}(|n\rangle \langle n|)=0$.

\item[b)] For the initial state
\begin{equation}
|\psi _{0}\rangle =\frac{|0\rangle +\alpha |1\rangle }{\sqrt{1+|\alpha |^{2}}%
},\quad |\alpha |\ll 1,  \label{sCS}
\end{equation}%
which approximately describes a low excited coherent state, the short time
solution (\ref{stE}) has the form
\begin{eqnarray}
\hat{R}(t) &\approx &\frac{1}{1+|\alpha |^{2}}\left( |0\rangle \langle
0|+|\alpha |^{2}|1\rangle \langle 1|+\alpha \left( 1-2it\right) |1\rangle
\langle 0|+\alpha ^{\ast }(1+2it)|0\rangle \langle 1|\right.  \notag \\
&&\left. -\frac{it}{\sqrt{2}}\left( \alpha |2\rangle \langle 1|-\alpha
^{\ast }|1\rangle \langle 2|\right) \right) +O(t^{2}),  \label{kerr1}
\end{eqnarray}%
where the second line describes the action of the Lindbland-like operator $%
\mathcal{L}_{Kerr}$, satisfying Eq.(\ref{OR1}). It is easy to see that for
short times there are three non-zero eigenvalues: $\lambda _{1}=1-O(t^{2})$,
\begin{equation}
\lambda _{\pm }=\pm \frac{|\alpha |t}{\sqrt{2}\left( 1+|\alpha |^{2}\right)
^{3/2}}+O(t^{2}),  \label{kerr2}
\end{equation}%
and the negative one appears with the rate $\lambda _{-}\sim |\alpha |/\sqrt{%
2}$. Interestingly, the same result can be obtained by optimizing the
min-max solution in the orthogonal to the state (\ref{sCS}) subspace. For
instance, considering a sample state%
\begin{equation*}
|\phi \rangle =\frac{\alpha ^{\ast }|0\rangle -|1\rangle +\beta |2\rangle }{%
\sqrt{1+|\alpha |^{2}+|\beta |^{2}}},\;\langle \phi |\psi _{0}\rangle =0,
\end{equation*}%
we immediately obtain that $\underset{\beta }{\min }\langle \phi |\hat{R}%
(t)|\phi \rangle =\lambda _{-}.$

\item[c)] For the initial coherent state $|\alpha \rangle $ we make use the
min-max theorem (\ref{min}) and for the sample state
\begin{equation}
|\phi \rangle =\frac{|\beta \rangle -\langle \alpha |\beta \rangle |\alpha
\rangle }{\sqrt{1+|\langle \beta |\alpha \rangle |^{2}}},\;\langle \alpha
|\phi \rangle =0,  \label{CSort}
\end{equation}%
one obtains the following upper bound for the negative eigenvalue of $\hat{R}%
(t)$ for $|\alpha |\gg 1$
\begin{equation*}
\lambda _{-}\leq \underset{\beta }{\min }\langle \phi |\hat{R}(t)|\phi
\rangle \approx -\frac{|\alpha |^{3}e^{-|\alpha |^{2}}t}{\sqrt{2e}}.
\end{equation*}
\end{description}

The fidelity (\ref{Fg}) also is deviated very slowly from the unity for the
initial coherent state $|\alpha \rangle $ in the scale of classical period
of oscillations $T_{cl}\sim |\alpha |^{-2}$,%
\begin{equation}
\mathcal{F}(t)\sim 1-\frac{3}{2}|\alpha |^{2}t^{2}=1-\frac{3}{2}|\alpha
|^{-2}\left( \frac{t}{T_{cl}}\right) ^{2},  \label{FK}
\end{equation}%
while the self-correlation function (\ref{Gt}) exhibits a fast change of the
initial state at short times,%
\begin{equation*}
G(t)\sim 1-4|\alpha |^{6}t^{2}=1-4|\alpha |^{2}\left( \frac{t}{T_{cl}}%
\right) ^{2}.
\end{equation*}

\subsection{Second harmonic generation}

The Hamiltonian describing the effect of second harmonic generation (and
down conversion) is of third order%
\begin{equation*}
\hat{H}_{SG}=\{\hat{a}^{\dagger 2}\hat{a}+\hat{a}^{\dagger }\hat{a}%
^{2}\}_{sym}.
\end{equation*}%
The effective Hamiltonian and the set of Lindblad operators are%
\begin{eqnarray}
\hat{H}_{eff} &=&\frac{3}{8}\left( \hat{a}^{\dagger 2}\hat{a}+\hat{a}\hat{a}%
^{\dagger 2}+\hat{a}^{\dagger }\hat{a}^{2}+\hat{a}^{2}\hat{a}^{\dagger
}\right) =\frac{3}{4}\hat{H}_{SG},  \label{HeffSG} \\
\mathcal{L} &=&\frac{1}{16}\left( \mathcal{L}_{01}^{21}+\tilde{\mathcal{L}}%
_{02}^{21}-\mathcal{L}_{20}^{21}-\tilde{\mathcal{L}}_{01}^{21}\right)
\label{LSG} \\
&&+\frac{1}{8}\left( \mathcal{L}_{10}^{21}+\tilde{\mathcal{L}}_{11}^{21}-%
\mathcal{L}_{11}^{21}-\tilde{\mathcal{L}}_{10}^{21}\right) ,  \notag
\end{eqnarray}%
where%
\begin{align*}
L_{01}^{21}& =\hat{a}-i\hat{a}^{2}, & \tilde{L}_{01}^{21}& =\hat{a}+i\hat{a}%
^{2}, \\
\tilde{L}_{02}^{21}& =i\left( \hat{a}^{\dagger }-i\hat{a}^{\dagger 2}\right)
, & L_{20}^{21}& =-i\left( \hat{a}^{\dagger }+i\hat{a}^{\dagger 2}\right) ,
\\
L_{10}^{21}& =\hat{a}^{\dagger }-i\hat{a}\hat{a}^{\dagger }, & \tilde{L}%
_{10}^{21}& =\hat{a}^{\dagger }+i\hat{a}\hat{a}^{\dagger }, \\
\tilde{L}_{11}^{21}& =i\left( \hat{a}-i\hat{a}^{\dagger }\hat{a}\right) , &
L_{11}^{21}& =-i\left( \hat{a}+i\hat{a}^{\dagger }\hat{a}\right) .
\end{align*}%
In the representation (\ref{LF}) the operator (\ref{LSG}) has the form
\begin{eqnarray*}
F(\hat{R}) &=&\frac{1}{8}(2\hat{a}\hat{R}\hat{a}^{\dagger 2}+2\hat{a}%
^{\dagger }\hat{R}\hat{a}^{2}+4\hat{a}\hat{R}\hat{a}^{\dagger }\hat{a}+4\hat{%
a}^{\dagger }\hat{R}\hat{a}\hat{a}^{\dagger } \\
&&+\hat{a}^{\dagger 2}\hat{a}\hat{R}+\hat{R}\hat{a}^{\dagger 2}\hat{a}+\hat{a%
}^{2}\hat{a}^{\dagger }\hat{R}+\hat{R}\hat{a}^{2}\hat{a}^{\dagger }).
\end{eqnarray*}%
The construction of the Kraus operators is similar to the previous example.

It is straightforward to find that for the vacuum initial state
\begin{equation*}
|\psi _{0}\rangle =|0\rangle ,
\end{equation*}%
the short-time expansion for $\hat{R}(t)$ has the form
\begin{equation*}
\hat{R}(t)\approx |0\rangle \langle 0|+\frac{it}{2}\left( |0\rangle \langle
1|-|1\rangle \langle 0|\right) +\frac{\sqrt{2}it}{4}\left( |1\rangle \langle
2|-|2\rangle \langle 1|\right) +O(t^{2}),
\end{equation*}%
leading to the following negative eigenvalue
\begin{equation}
\lambda _{-}=-\frac{t}{2\sqrt{2}}+O(t^{2}).  \label{second1}
\end{equation}%
For the coherent state $|\alpha \rangle $ we proceed as in the previous
example of Kerr Hamiltonian, minimizing the average value of (\ref{stE})
over the sample states (\ref{CSort}) obtaining for $|\alpha |\gg 1$,
\begin{equation*}
\lambda _{-}\leq -\frac{|\alpha |^{2}e^{-|\alpha |^{2}}}{2\sqrt{2e}}t.
\end{equation*}%
The fidelity (\ref{Fg}) behaves very differently for the initial coherent
and number states. In particular, for a number state $|N\rangle $ one obtains%
\begin{equation*}
1-\mathcal{F}(t)\sim \frac{1}{8}\left( 10N^{3}+6N^{2}+10N+3\right) t^{2},
\end{equation*}%
while for coherent states $|\alpha \rangle $,%
\begin{equation*}
1-\mathcal{F}(t)\sim \frac{3}{8}t^{2}.
\end{equation*}%
This confirms the intuition that TWA works significantly better for coherent
states than for number states with the same average energy.

In Conclusion we have shown that the Truncated Wigner Approximation (\ref%
{twa eq1}) developed in phase-space corresponds to a Lindblad-type
(non-unitary) evolution with indefinite metric in the space of linear
operators, except for the case of linear evolution (governed by quadratic
Hamiltonians). As a result, the inverse image of the classically evolved
Wigner function is a non-positive operator $\hat{R}(t)$ satisfying the
relations (\ref{R1})-(\ref{R2}). In other words, the classical dynamics
generates a Kraus-like map containing both positive and negative terms that
transform an initial density matrix into an operator not corresponding to a
physical state. The positivity of the operator corresponding to the
classically evolved Wigner function is broken for $t=0^{+}$ in case of
anharmonic evolution. Observe, that the unitarity is lost by the
approximation
\begin{equation*}
\lbrack \hat{w}(\alpha ),\hat{H}]\rightarrow \{\hat{w}(\alpha ),W_{H}\},
\end{equation*}%
which should be considered as a weak asymptotic limit on the states with $%
\bar{n}=Tr(\hat{a}^{\dagger }\hat{a}\hat{\rho})\rightarrow \infty $ (which
basically corresponds to considering only a zero-order term of singularly
perturbed phase-space evolution equation).

It is interesting to contrast the quantum and TWA dynamics of an initial
coherent state, described in the phase-space by the Wigner function Eq.(\ref%
{WG}):

a) the quantum and TWA harmonic evolutions preserve the positivity of the
Wigner function and positive definiteness of its inverse image, which in
this case coincides with the density matrix of the evolved state;

b) during an anharmonic TWA evolution the positivity of the Wigner function
is maintained, but the operator counterpart of $W_{\rho }(\alpha (-t))$\ is
not a positively defined operator; the quantum anharmonic evolution in
phase-space leads to a non-positive Wigner function, the negativity of which
can be used for the detection of non-classicality \cite{KZ}.

This work is partially supported by the Grant 254127 of CONACyT (Mexico).

\appendix

\section{}
Here we show that the Liuovillian evolution in phase-space does not
correspond to quantum unitary dynamics.

The classical phase-space evolution of the Wigner function can be
symbolically described as%
\begin{equation*}
W_{\rho }\left( t\right) =e^{t\left\{ W_{H},\cdot \right\} }W_{\rho }\left(
0\right) ,
\end{equation*}%
where $W_{H}(\alpha ,\alpha ^{\ast })$ is the symbol of the Hamiltonian and
the flat space Poisson brackets has the form%
\begin{equation*}
\left\{ \cdot ,\cdot \right\} =i\left( \partial _{\alpha }\otimes \partial
_{\alpha ^{\ast }}-\partial _{\alpha ^{\ast }}\otimes \partial _{\alpha
}\right) ,
\end{equation*}%
the symbol $\otimes $ indicates the order of application of the derivatives.
The TWA\ evolution of the square of the density operator can be determined
in terms of the star-product as follows
\begin{equation*}
W_{\rho ^{2}}\left( t\right) =e^{-\frac{i}{2}\left\{ \cdot ,\cdot \right\}
}\left( W_{\rho }\left( t\right) W_{\rho }\left( t\right) \right) =W_{\rho
}\left( t\right) \ast W_{\rho }\left( t\right) ,
\end{equation*}%
where $\ast $ denotes the star-product operator.

Since $\left\{ W_{H},\cdot \right\} $ is a first-order differential operator
we can write%
\begin{equation}
e^{t\left\{ W_{H},\cdot \right\} }\left( W_{\rho }\left( 0\right) \ast
W_{\rho }\left( 0\right) \right) =\hat{S}_{t}\left( W_{\rho }\left( \alpha
(-t\right) )W_{\rho }\left( \alpha (-t\right) )\right) ,  \label{SW}
\end{equation}%
where
\begin{equation*}
W_{\rho }\left( \alpha (-t\right) )=e^{t\left\{ W_{H},\cdot \right\}
}W_{\rho }(0),
\end{equation*}%
is the classically evolved Wigner function (\ref{Wt}) and
\begin{equation}
\hat{S}_{t}=\exp \left( \frac{1}{2}e^{t\left\{ W_{H},\cdot \right\} }\left(
\partial _{\alpha }\otimes \partial _{\alpha ^{\ast }}-\partial _{\alpha
^{\ast }}\otimes \partial _{\alpha }\right) e^{-t\left\{ W_{H},\cdot
\right\} }\right) ,  \label{St}
\end{equation}%
is the transformed star-product operator. \ For quadratic in $\alpha $ and $%
\alpha ^{\ast }$ symbols $W_{H}(\alpha ,\alpha ^{\ast })$ the operator (\ref%
{St}) is invariant under transformation generated by $W_{H}$. This leads to
the well known result,%
\begin{equation}
W_{\rho }\left( \alpha (-t\right) )\ast W_{\rho }\left( \alpha (-t\right)
)=W_{\rho }\left( \alpha (-t\right) ),  \label{WW}
\end{equation}%
i.e. an initial pure state evolves into a pure state under action of
quadratic Hamiltonians.

In general case, by taking into account that%
\begin{eqnarray*}
e^{t\left\{ H,\cdot \right\} }\partial _{\alpha }e^{-t\left\{ H,\cdot
\right\} } &=&\partial _{\alpha }-it\left( \partial _{\alpha
}^{2}W_{H}\partial _{\alpha ^{\ast }}-\partial _{\alpha }\partial _{\alpha
^{\ast }}W_{H}\partial _{\alpha }\right) +O(t^{2}), \\
e^{t\left\{ H,\cdot \right\} }\partial _{\alpha ^{\ast }}e^{-t\left\{
H,\cdot \right\} } &=&\partial _{\alpha ^{\ast }}+it\left( \partial _{\alpha
^{\ast }}^{2}W_{H}\partial _{\alpha }-\partial _{\alpha }\partial _{\alpha
^{\ast }}W_{H}\partial _{\alpha ^{\ast }}\right) +O(t^{2}),
\end{eqnarray*}%
we arrive at the following deformation of the star-product operator at short
times,%
\begin{equation*}
\hat{S}=\exp \left[ -\frac{i}{2}\left\{ \cdot ,\cdot \right\} +\frac{t}{2}%
\hat{d}\right] ,
\end{equation*}%
being
\begin{equation}
\hat{d}=\partial _{\alpha }\otimes \{\partial _{\alpha ^{\ast }}W_{H},\cdot
\}-\{\partial _{\alpha ^{\ast }}W_{H},\cdot \}\otimes \partial _{\alpha
}+\{\partial _{\alpha }W_{H},\cdot \}\otimes \partial _{\alpha ^{\ast
}}-\partial _{\alpha ^{\ast }}\otimes \{\partial _{\alpha }W_{H},\cdot \},
\label{d}
\end{equation}%
the first-order defect operator. It is easy to see that for non-harmonic
Hamiltonians the defect becomes non-trivial.

For instance, in the case of Kerr evolution, $W_{H}\sim \left\vert \alpha
\right\vert ^{4}$,
\begin{equation*}
\hat{d}\sim \partial _{\alpha }\otimes \alpha ^{2}\partial _{\alpha }-\alpha
^{2}\partial _{\alpha }\otimes \partial _{\alpha }+c.c.+2\left\vert \alpha
\right\vert ^{2}\partial _{\alpha }\otimes \partial _{\alpha ^{\ast
}}-\partial _{\alpha }\otimes 2\left\vert \alpha \right\vert ^{2}\partial
_{\alpha ^{\ast }}+c.c.
\end{equation*}%
Thus, in general the relation (\ref{WW}) is not fulfilled, so that
\begin{equation}
\hat{R}(t)\neq \hat{R}^{2}(t),  \label{RR}
\end{equation}%
i.e. the initial pure state becomes mixed except when

a) $\left\{ W_{H},W_{\rho }(0)\right\} =0,$ i.e. the initial state is an
eigenstate of the Hamiltonian;

b) The Hamiltonian is a quadratic function in $a$ and $a^{\dagger }$.

The above result can be also obtained directly from the evolution equation (%
\ref{Req}), by observing that the dynamics of $\hat{R}^{2}(t)$ is described
by
\begin{equation*}
\partial _{t}\hat{R}^{2}=i[\hat{R}^{2},\hat{H}_{eff}]+\hat{R}\mathcal{L}(%
\hat{R})+\mathcal{L}(\hat{R})\hat{R},
\end{equation*}%
and for an initial pure state $\hat{R}^{2}(0)=\hat{R}(0)=|\psi \rangle
\langle \psi |$. However, due to the structure of the non-Hamiltonian part (%
\ref{LF}), $\mathcal{L}(\hat{R})=i\left( F(\hat{R})-F^{\dag }(\hat{R}%
)\right) $,
\begin{equation}
\mathcal{L}(\hat{R})\neq \hat{R}\mathcal{L}(\hat{R})+\mathcal{L}(\hat{R})%
\hat{R},  \label{LR}
\end{equation}%
unless $F(\hat{R})=F^{\dag }(\hat{R})$ or $\mathcal{L}(\hat{R})=0$. Thus,
the evolution of $\hat{R}(t)$ and $\hat{R}^{2}(t)$ is completely different.

\section{}

In this Appendix we outline the derivation of Eq.(\ref{Req}-\ref{lin3}).

Applying the first-order differential operator in the right-hand side of (%
\ref{EqW}) to the kernel (\ref{w}) and making use the correspondence rules%
\begin{eqnarray*}
\partial _{\alpha }\hat{w}(\alpha ) &=&\hat{a}^{\dagger }\hat{w}(\alpha )-%
\hat{w}(\alpha )\hat{a}^{\dagger }, \\
\alpha \hat{w}(\alpha ) &=&\frac{\hat{a}\hat{w}(\alpha )+\hat{w}(\alpha )%
\hat{a}}{2},
\end{eqnarray*}%
and their generalizations%
\begin{eqnarray*}
\alpha ^{k}W(\alpha ) &=&\frac{1}{2^{k}}Tr\left( \sum_{l=0}^{k}C_{k}^{l}\hat{%
a}^{k-l}\hat{\rho}\hat{a}^{l}\hat{w}(\alpha )\right) , \\
\alpha ^{\ast k}W(\alpha ) &=&\frac{1}{2^{k}}Tr\left( \sum_{l=0}^{k}C_{k}^{l}%
\hat{a}^{\dagger k-l}\hat{\rho}\hat{a}^{\dagger l}\hat{w}(\alpha )\right) ,
\end{eqnarray*}%
we arrive at Eq.(\ref{Req}-\ref{lin3}) after the normal re-ordering
\begin{equation*}
\hat{a}^{k}\hat{a}^{\dagger l}=\sum_{p=0}^{\min (k,l)}\frac{k!l!}{%
(k-p)!(l-p)!}\hat{a}^{\dagger l-p}\hat{a}^{k-p}.
\end{equation*}

\section{}

Here we give explicit expressions of the Hamiltonians (\ref{Hq}) and (\ref%
{Heff}) in the normal ordered form,
\begin{eqnarray*}
H_{eff}=\frac{n+m}{2^{n+m}} &&\left( 2a^{\dag
m}a^{n}+\sum_{k=1}^{m}C_{m}^{k}C_{n}^{k}k!a^{\dag m-k}a^{n-k}\right. \\
&&\left. +2a^{\dag n}a^{m}+\sum_{k=1}^{m}C_{m}^{k}C_{n}^{k}k!a^{\dag
n-k}a^{m-k}\right) ,
\end{eqnarray*}%
and
\begin{eqnarray*}
H_{mn} &=&a^{\dag m}a^{n}+\sum_{k=1}^{m}C_{m}^{k}C_{n}^{k}k!\left( \frac{1}{2%
}\right) ^{k}a^{\dag m-k}a^{n-k} \\
&&+a^{\dag n}a^{m}+\sum_{k=1}^{m}C_{m}^{k}C_{n}^{k}k!\left( \frac{1}{2}%
\right) ^{k}a^{\dag n-k}a^{m-k},
\end{eqnarray*}%
here $n\geq m$. One can observe that $H_{eff}\sim H_{mn}$ only for $m=1$ and
$H_{eff}=\frac{1}{2}H_{mn}+\frac{1}{2}$ in the particular case $n=m$ $=2$.


\begin{thebibliography}{99}
\bibitem{wf} The Wigner function in the flat phase-space can be also defined
in an equivalent way in terms of position and momentum coordinates, $W_{\rho
}(q,p)=\int dq^{\prime }\langle q+q^{\prime }/2|\hat{\rho}|q-q^{\prime
}/2\rangle \exp \left( -ipq^{\prime }/\hbar \right) $, where $q=\text{Re}%
\alpha /\sqrt{2}$, $p=\text{Im}\alpha /\sqrt{2}$.
\bibitem{H} An arbitrary Hamiltonian (with real coefficients) on $a$ and $%
a^{\dagger }$ can be expanded as $H=\sum_{mn}c_{mn}a^{\dagger m}a^{n}+h.c.$.
Monomials $a^{\dagger m}a^{n}$ can be always expressed as a linear
combination of $\{a^{\dagger m-k}a^{n-k}+a^{\dagger n-k}a^{m-k}\}_{sym}$, $%
k=0,...,\min (m,n)$.
\end{thebibliography}
\end{document}